\documentclass[twocolumn,preprintnumbers, 
               superscriptaddress,
               eqsecnum,nofootinbib]{revtex4-1}
\usepackage{graphicx, fancybox}
\usepackage{amsmath,amssymb}
\usepackage[colorlinks=true, pdfstartview=FitV, linkcolor=red, citecolor=blue, urlcolor=blue]{hyperref}
\usepackage{color}

\newcommand{\vect}[1]{\mathbf{#1}}

\newcommand{\vp}{\vect{p}}

\newcommand{\vzero}{\vect{0}}

\newcommand{\vB}{\vect{B}}

\newcommand{\Bf}{B_f}

\newcommand{\comment}[1]{}

\newcommand{\Nc}{N_c}

\usepackage{color}

\begin{document}

\title{
Gluon Spectrum in Quark-Gluon Plasma under Strong Magnetic Fields
}

\author{Koichi Hattori}
\email{koichi.hattori@riken.jp}
\affiliation{Physics Department and Center for Particle Physics and Field Theory, 
Fudan University, Shanghai 200433, China}

\author{Daisuke Satow}
\email{dsato@th.physik.uni-frankfurt.de}
\affiliation{Goethe University Frankfurt am Main, Institute for Theoretical Physics, Max-von-Laue-Str. 1, D-60438 Frankfurt am Main, Germany}

\begin{abstract} 
We obtain the %most 
general expression of the gluon propagator at finite temperature ($T$) 
and in a magnetic field ($B$), 
for the case that the four transverse tensor structures appear in the gluon self-energy. 
By using this expression and a specific form of the one-loop gluon self-energy in the lowest Landau level approximation, 
we analyze the gluon spectrum in the strong magnetic field limit.
As a result, we find that there exist two collective excitations 
of which the energies are of the order of $p\sim gT$ with $g$ being the coupling constant.
One of the two excitations enjoys properties quite different from 
those of the collective excitations at $B=0$ which have been discussed by using the hard thermal loop approximation.
We also discuss the static and dynamical screening effects, 
which are expected to be important for computation of transport coefficients in strong magnetic fields.
\end{abstract} 

\date{\today}

\maketitle

%%%%%%%%%%%%%%%%%%%%%%%%%%%%%%%%%%%%%%%%%%%
\section{Introduction}
\label{sec:intro}

The extreme state of matter created by the relativistic heavy-ion collisions 
provides us with an opportunity to investigate 
the dynamics governed by quantum chromodynamics (QCD) at high temperature ($T$). 
Motivated by the experiments, a number of theoretical studies has addressed 
the properties of the extreme QCD matter called the quark-gluon plasma (QGP). 
Recent studies, both in theory and experiment, also suggest opportunities 
of studying novel properties of the QGP in the strong magnetic field ($B$) 
which is thought to be induced in the heavy-ion collisions~\cite{
Skokov:2009qp, Voronyuk:2011jd, Bzdak:2011yy, Deng:2012pc, Deng:2014uja} 
(see Refs.~\cite{Huang:2015oca, Hattori:2016emy} for recent reviews on the estimates of the strengths). 
The outcomes of these studies suggest not only 
the emergence of the nondissipative transport phenomena~\cite{Kharzeev:2007jp, Fukushima:2008xe} 
(see also Refs.~\cite{Kharzeev:2015znc, Skokov:2016yrj, Huang:2015oca, Hattori:2016emy} for recent reviews), 
but also the drastic changes in the conventional transport phenomena such as 
the shear viscosity~\cite{Tuchin:2011jw}, the heavy-quark diffusion dynamics~\cite{Fukushima:2015wck}, 
the jet energy loss~\cite{Li:2016bbh}, and the electrical conductivity~\cite{
Hattori:2016lqx, Hattori:2016cnt, Satow:2014lia, Kerbikov:2014ofa, Nam:2012sg, Gorbar:2016qfh, Pu:2014fva, Buividovich:2010tn}. 
The quark spectrum, which is one of the fundamental building blocks of QGP, 
was also suggested to show a drastic change~\cite{Elmfors:1996fx}.
Not only these intrinsic properties of the QGP, but also the macroscopic time evolution of the QGP in the heavy ion collision was investigated on the basis of the anomalous hydrodynamics~\cite{Hongo:2013cqa, Hirono:2014oda, Jiang:2016wve} and 
the magnetohydrodynamics~\cite{Pu:2016ayh, Roy:2015kma, Inghirami:2016iru, Pu:2016bxy}.

In this paper, we discuss the gluon spectrum in the strong magnetic field. 
In the case without the magnetic field, the preceding studies have clearly shown 
that investigating the properties of the gluon, a fundamental degree of freedom in the QGP, 
is important for understanding many aspects of the QGP 
from basic excitations to more complex phenomena. 
One important example is the computation of transport coefficients:
In the 2-to-2 scattering process that appears in the leading-order calculation of 
the transport coefficients at $B=0$~\cite{Arnold:2000dr, Arnold:2003zc, Gagnon:2007qt, Gagnon:2006hi}, 
the exchanged gluon has a small energy/momentum compared with the temperature. 
Therefore, the Debye and dynamical screening properties of this soft gluon 
is the necessary ingredient. 
The whole task is carried out by the computation of the gluon self-energy 
and the resummation procedure called the hard thermal loop (HTL) 
resummation~\cite{Pisarski:1988vd, Braaten:1989kk, Braaten:1989mz, Braaten:1990it, Braaten:1992gd, Kobes:1992ys}.

%%%%%%%%%%%%%%%%%%%%

We investigate the general expression of the gluon propagator at finite temperature and in a magnetic field,  assuming the four transverse tensor components in the gluon self-energy that have been known to appear in the perturbative calculation~\cite{Hattori:2012je,Pisarski:1988vd, Braaten:1989kk, Braaten:1989mz, Braaten:1990it, Braaten:1992gd, Kobes:1992ys}. 
Then, we analyze the gluon spectrum in the momentum region of the order of $gT$, with $g$ being the QCD coupling constant, 
by using a specific gluon self-energy at the one-loop order in the lowest Landau level (LLL) approximation. 
We find a novel collective excitation in this momentum region 
and closely look into the dispersion relation and the strength to uncover its basic properties. 
We also investigate the Debye and dynamical screening effects 
to provide the gluon propagator which
will be important for, e.g., 
the computation of the transport coefficients in the future studies. 

We note that, when the magnetic field is so strong that the LLL 
approximation is reliable, 
it has been known that one of the two physical modes of the gluon 
is screened by the interaction effect with the quarks in the LLL~\cite{Miransky:2002rp,*Fukushima:2011nu}. 
Moreover, the screening mass has no temperature dependence 
since this mass is interpreted as the Schwinger mass 
originated from the dimensional reduction in the LLL~\cite{Schwinger:1962tp}.
However, the other mode, which is not screened by the quarks in the LLL, has not been studied in detail. 
Our analysis sheds light on this point.

%%%%%%%%%%%%%%%%%%%%
This paper is organized as follows:
In the next section, we obtain the general form of the gluon propagator at finite $T$ and $B$ in the covariant gauge 
by %using the general tensor structure of the gluon self-energy determined from the symmetries.
 assuming the tensor structure of the gluon self-energy suggested by the leading-order calculation~\cite{Hattori:2012je,Pisarski:1988vd, Braaten:1989kk, Braaten:1989mz, Braaten:1990it, Braaten:1992gd, Kobes:1992ys}.
We also discuss the physical meaning of the excitations obtained from this propagator 
with the use of the explicit forms of the polarization vectors. 
In Sec.~\ref{sec:one-loop}, we use the specific expression of the self-energy at the one-loop order 
in the strong magnetic field, and investigate the spectrum of the gluon excitation in a few energy scales. 
We also discuss the static and the dynamical screening effect. 
Section~\ref{sec:summary} is devoted to the summary of this paper. 
We evaluate the gluon propagator in the Coulomb gauge in Appendix~\ref{app:Coulomb}, 
and summarize the properties of the projection tensors 
used in order to evaluate the gluon propagator in Appendix~\ref{app:projection}.

%%%%%%%%%%%%%%%%%%%%%%%%%%%%%%%%%%%%%%%%%%%
\section{General form of Gluon Propagator}
\label{sec:general}

In this section, we first show the %most 
general form of the gluon propagator 
with the general gluon momentum and %all possible tensor structures of the gluon self-energy. 
the tensor structures of the gluon self-energy suggested by the leading-order calculation~\cite{Hattori:2012je,Pisarski:1988vd, Braaten:1989kk, Braaten:1989mz, Braaten:1990it, Braaten:1992gd, Kobes:1992ys}. 
We then discuss the specific momentum configurations to clarify the physical picture. 

\subsection{General momentum}
\label{ssc:gluon-general}

We consider the retarded gluon propagator $D^R_{\mu\nu}(p)$ 
of which the color indices are suppressed for the notational simplicity. 
This propagator is related to the retarded gluon self-energy ($\varPi^R$) as 
\begin{align}
\label{eq:D-D0Pi}
D^R_{\mu\nu}(p)
&= [(D^0(p))^{-1}+\varPi^R(p)]^{-1}_{\mu\nu},
\end{align}
where $D^0$ is the bare propagator.
Here, the inverse matrix in the Minkowski space is defined as $D^{\mu\nu} D^{-1}_{\nu\alpha}=g^\mu_\alpha$. 
We adopt the covariant gauge to get a simpler tensor structure than in the other gauges.
For some purposes such as the computation of transport coefficients, 
the ghost-free gauges are more convenient. 
Therefore, we give the propagator in the Coulomb gauge in Appendix~\ref{app:Coulomb}.

The bare propagator in the covariant gauge reads
\begin{align}
\label{eq:D-free}
D^0_{\mu\nu}(p)
&= -\frac{P^0_{\mu\nu}(p)}{p^2}+\alpha\frac{p_\mu p_\nu}{(p^2)^2},
\end{align}
where $\alpha$ is the gauge-fixing parameter and 
$P^0_{\mu\nu}(p)\equiv -(g_{\mu\nu}-p_\mu p_\nu/p^2)$ is the projection tensor
into the transverse component in the Lorentz-symmetric system. 
We note that the gluon energy $p^0$ appearing in the above expression 
contains an infinitesimal imaginary part ($p^0+i\epsilon$) for the retarded function.
The inverse matrix is 
\begin{align}
[D^0]^{-1}_{\mu\nu}(p)
&= -p^2 P^0_{\mu\nu}(p)+\frac{1}{\alpha}p_\mu p_\nu.
\end{align}

Now, we look at the self-energy terms.
Let us consider what kind of tensor structures generally appear at finite $T$ and $B$.
One can set the direction of the magnetic field along the $z$-axis without losing generality.
Then, the independent tensors one can use to construct $\varPi^{\mu \nu}_R(p)$ are 
\begin{align}
\begin{split}
& p^\mu p^\nu, n^\mu n^\nu, b^\mu b^\nu, \\
&(p^\mu n^\nu+n^\mu p^\nu), (p^\mu b^\nu+b^\mu p^\nu), (b^\mu n^\nu+n^\mu b^\nu), 
g^{\mu\nu},
\end{split}
\end{align}
where $n^\mu\equiv (1,\vzero)$ and $b^\mu\equiv (0,0,0,-1)$ break 
the Lorentz and rotational symmetries, respectively. 
The latter vector indicates the preferred direction in the presence of the magnetic field. 
Instead of the last four tensors, it is convenient to use the four projection tensors, 
\begin{align} 
P^{\mu\nu}_T(p)
&= -g^{\mu\nu} + \frac{p^0}{\vp^2} (p^\mu n^\nu+ n^\mu p^\nu)
\nonumber\\
&~~~  
-\frac{1}{\vp^2}\left( p^\mu p^\nu  + p^2 n^\mu n^\nu\right),
\label{eq:PT-def}\\
\nonumber
P^{\mu\nu}_L(p)
&=-\frac{p^0}{\vp^2}(p^\mu n^\nu+ n^\mu p^\nu)
\nonumber\\
&~~~~~ +\frac{1}{\vp^2} \left[ \frac{(p^0)^2}{p^2} p^\mu p^\nu +p^2n^\mu n^\nu \right], 
\\
P^{\mu\nu}_\parallel(p)
&= - \frac{p^0 p^3}{p^2_\parallel} (b^\mu n^\nu+n^\mu b^\nu)
\nonumber\\
&~~~~~ + \frac{1}{p^2_\parallel}\left[   (p^0)^2b^\mu b^\nu +(p^3)^2n^\mu n^\nu \right]
\\
&= -\left(g^{\mu\nu}_\parallel-\frac{p^{\mu}_\parallel p^{\nu}_\parallel }{p^2_\parallel}\right)
,\\
\nonumber
P^{\mu\nu}_\perp(p)
&= \frac{1}{\vp^2_\perp}  [ -\vp^2_\perp g^{\mu\nu} + p^0 (p^\mu n^\nu+n^\mu p^\nu) 
\\
&~~~ - p^3 (p^\mu b^\nu+b^\mu p^\nu) + p^0p^3 (b^\mu n^\nu+n^\mu b^\nu)
\nonumber\\
&~~~ -p^\mu p^\nu + \left(\vp^2_\perp-(p^0)^2\right) n^\mu n^\nu  - \vp^2 b^\mu b^\nu ]
\\
\label{eq:Pperp-def}
&= -\left(g^{\mu\nu}_\perp-\frac{p^{\mu}_\perp p^{\nu}_\perp }{p^2_\perp}\right).
\end{align}
We have defined $g^\parallel_{\mu\nu}=(1,0,0,-1)$, $g^\perp_{\mu\nu}=(0,-1,-1,0)$, 
$p^\mu_\parallel = g^{\mu\nu}_\parallel p_\nu$, 
$p^\mu_\perp = g^{\mu\nu}_\perp p_\nu$, 
$p^2_\parallel = (p^0)^2-(p^3)^2$, and $\vp^2_\perp = (p^1)^2+(p^2)^2$.
All four of the above projection tensors are transverse to the momentum as $p_\mu P^{\mu\nu}_i(p)=0$. 
We note that, the former two tensors ($P_T, P_L$) are known to appear at finite $T$ and $B=0$ case~\cite{Pisarski:1988vd, Braaten:1989kk, Braaten:1989mz, Braaten:1990it, Braaten:1992gd, Kobes:1992ys}, while the latter two ($P_\parallel, P_\perp$) appear at $T=0$ and finite $B$ case~\cite{Hattori:2012je}, by the perturbative calculations.

Then, the tensor structure of the self-energy can be written as~\cite{Bordag:2008wp} 
\begin{align}
\label{eq:selfenergy}
\begin{split}
\varPi^{\mu \nu}_R(p)
&= \sum_{i=T, L, \parallel, \perp}\varPi_i(p) P^{\mu\nu}_i(p) \\
&~~~+ \varPi_p\frac{p^\mu p^\nu}{p^2}
+\varPi_n n^\mu n^\nu
+\varPi_b b^\mu b^\nu.
\end{split}
\end{align}
We note that one cannot make other transverse tensors 
that are independent of the above four projection tensors, 
by using $p^\mu p^\nu$, $n^\mu n^\nu$, and $b^\mu b^\nu$.
%The gluon self-energy satisfies the Ward identity for the non-Abelian theory derived from the gauge invariance of T-matrix~\cite{Braaten:1989mz}, $p_\mu p_\nu \varPi^{\mu \nu}_R(p)=0$, giving a constraint on the coefficients as $0=\varPi_p p^2+\varPi_n (p^0)^2+\varPi_b (p^3)^2$.
This is the most general form of the gluon self-energy at finite $T$ and $B$.
In the current paper, we only consider the terms that are 
proportional to the four transverse projection tensors, 
because only these four tensors appear in the leading-order perturbative calculation.
%the other terms do not appear in the final form of $D^R_{\mu\nu}$~\cite{Braaten:1990it}. 
%The explicit form of the self-energy at the one-loop order satisfies this assumption as we will see in the next section. 

%%%%%%%%%%%%%%%%%%%%%%%%%%%%%%%%%%%%%%%%%%%

\begin{figure*}[t] 
\begin{center}
\includegraphics[width=0.7\textwidth]{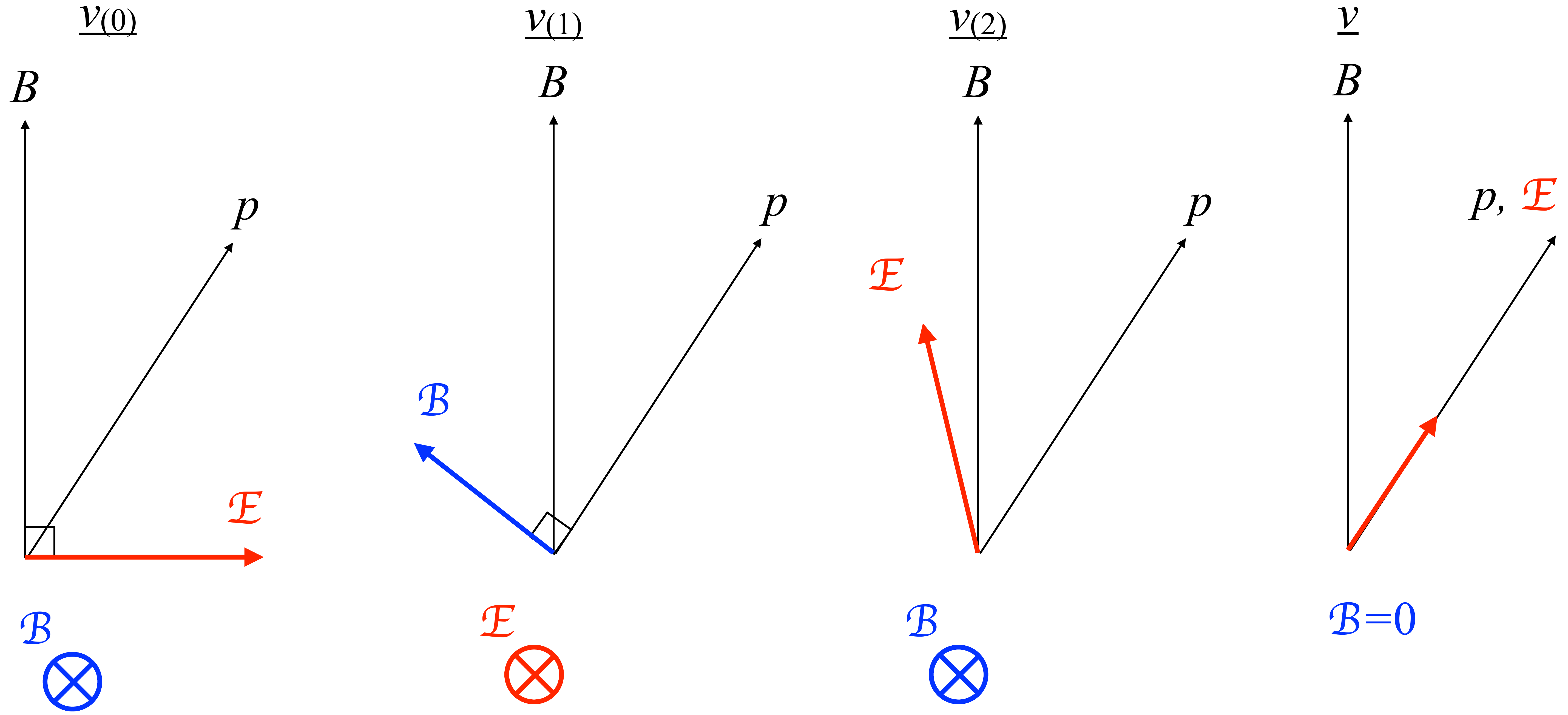} 
\caption{The directions of $\cal E$ and $\cal B$ generated by the four polarization vectors.
} 
\label{fig:polarization-generalp} 
\end{center} 
\end{figure*}

%%%%%%%%%%%%%%%%%%%%%%%%%%%%%%%%%%%%%%%%%%%

Now that the %general 
form of the self-energy is specified, 
one can obtain the corresponding gluon propagator 
by evaluating the inverse matrix appearing on the right-hand side of Eq.~(\ref{eq:D-D0Pi}).
By using the multiplicative properties of the four projection tensors 
summarized in Appendix~\ref{app:projection}, we find the general form of the gluon propagator as 
\begin{align} 
\label{eq:Dmunu-general}
\begin{split}
D^R_{\mu\nu}(p)
&= -\frac{1}{\Delta}
\Bigl[(p^2-\varPi_\parallel-\varPi_L)P^T_{\mu\nu}(p) \\
&~~~ +(p^2-\varPi_\parallel-\varPi_T)P^L_{\mu\nu}(p)
+\varPi_\parallel P^\parallel_{\mu\nu}(p) \\
&~~~ +D_\perp(p)P^\perp_{\mu\nu}(p)
\Bigr]
+\alpha\frac{p_\mu p_\nu}{(p^2)^2},
\end{split}
\end{align} 
where
\begin{align}
\nonumber 
\Delta&\equiv 
(p^2-\varPi_T)(p^2-\varPi_L) \\
\label{eq:Delta-general}
&~~~ -\varPi_\parallel\left[ p^2-\varPi_T a \frac{p^2}{p^2_\parallel}-\varPi_L(1-a)\frac{(p^0)^2}{p^2_\parallel }
\right], \\
\nonumber 
D_\perp(p) &\equiv 
\frac{1}{p^2-\varPi_T-\varPi_\perp} 
\Bigl[\varPi_\parallel(\varPi_L-\varPi_T)(1-a)\frac{(p^0)^2}{p^2_\parallel} \\
\label{eq:Dperp-general}
&~~~+\varPi_\perp(p^2-\varPi_L-\varPi_\parallel)\Bigr] ,
\end{align}
and $a\equiv (p^3)^2/\vp^2$.
We are not aware of the literature that has the general form of the gluon propagator at finite $T$ and $B$, 
 with the four tensor components in its self-energy.
We note that the gauge-fixing term is not affected by the self-energy at all.
Since this term does not reflect any physical property, we will not explicitly write this term below.

For later convenience, we introduce the three independent vectors as in Ref.~\cite{Hattori:2012je}: 
\begin{subequations}
\begin{align}
v_\mu^{(0)} &\equiv -\frac{1}{\sqrt{p^2p^2_\parallel\vp^2_\perp}}
\left(\vp^2_\perp p_0, p^2_\parallel p_1, p^2_\parallel p_2, \vp^2_\perp p_3 \right),\\
v_\mu^{(1)} &\equiv \frac{1}{\sqrt{\vp^2_\perp}}
\left(0, p_2, -p_1, 0\right),\\
v_\mu^{(2)} &\equiv - \frac{1}{\sqrt{p^2_\parallel}}
\left(p_3, 0, 0, p_0\right),
\end{align}
\end{subequations}
where $ p_0=p^0 $ and $ \vp_i = - \vp^i $ are understood. 
It is easy to show the orthogonality $v^\mu_{(i)} v_{(j)\mu}=0$ for $i\neq j$, 
the normalization, $v^\mu_{(i)} v_{(i) \mu}= -1$, and the transversality $p_\mu v^\mu_{(i)} =0$. 
In terms of these vectors, 
one can rewrite the projection tensors as $P^{\mu\nu}_\perp= v^\mu_{(1)} v^\nu_{(1)}$, 
$P^{\mu\nu}_\parallel=v^\mu_{(2)} v^\nu_{(2)}$, and $P^{\mu\nu}_0-P^{\mu\nu}_\perp-P^{\mu\nu}_\parallel= v^\mu_{(0)} v^\nu_{(0)}$.
One also finds $P^{\mu\nu}_L=v^\mu v^\nu$ where 
\begin{align}
\begin{split}
v^\mu &\equiv \frac{1}{|\vp|\sqrt{p^2}} 
\left[p^0 p^\mu-p^2 n^\mu \right] \\
&=\frac{1}{|\vp|\sqrt{p^2_\parallel}}
\left[-p^0|\vp_\perp|v^\mu_{(0)}+p^3\sqrt{p^2}v^\mu_{(2)} \right],
\end{split}
\end{align}
and then $P^{\mu\nu}_T=\sum_{i=0,1,2} v^\mu_{(i)} v^\nu_{(i)}-v^\mu v^\nu$.
$v^\mu$ is a superposition of $v^\mu_{(0)}$ and $v^\mu_{(2)}$, 
and does not have $v^\mu_{(1)}$ component.
Especially, we have $v^\mu\propto v^\mu_{(0)}$ when $\vp\perp\vB$, 
and $v^\mu \propto v^\mu_{(2)}$ when $\vp\parallel\vB$.

Physically, these vectors are the polarization vectors of the real-gluon field $A_\mu$. 
Properties of the induced color-electric ($\cal E$) and -magnetic ($\cal B$) fields\footnote{
In contrast to the case of the photon field~\cite{Hattori:2012je}, 
the field strength in QCD has the nonlinear terms in $A_\mu$ as well as the linear terms. 
These terms are expected to be negligible when the amplitude of $A^\mu$ is small, or the coupling constant $g$ is small. 
We focus on such cases and do not consider the nonlinear terms.} 
are also discussed in Appendix D of Ref.~\cite{Hattori:2012je}. 
We summarize the results in Fig.~\ref{fig:polarization-generalp}. 
In the all three modes $v^\mu_{(0)}$, $v^\mu_{(1)}$, and $v^\mu_{(2)}$, 
the induced $\cal E$ and $\cal B$ are orthogonal to each other, ${\cal E} \cdot{ \cal B}=0$. 
In the $v^\mu_{(0)}$ mode, the electric field lies in the plane spanned by $\vp$ and $B$, 
and the magnetic field extends in the out-of-plane direction. 
Notably, $\cal E$ is always orthogonal to the external $B$. 
In the $v^\mu_{(1)}$ mode, $\cal B$ lies in the $\vp$-$B$ plane, 
and $\cal E$ extends in the out-of-plane direction. 
In the $v^\mu_{(2)}$ mode, $\cal E$ lies in the $\vp$-$B$ plane
and $\cal B$ extends in the out-of-plane direction.
We note that only $v^\mu_{(2)}$ induces $\cal E$ that is not orthogonal to the external $B$.
This is the reason why the quarks at the LLL generates 
only the component $\varPi_\parallel$ in the gluon self-energy, which corresponds to $v^\mu_{(2)}$: 
The quarks at the LLL can move only in the direction of $B$, 
and thus are affected by the gluon excitation only when $\cal E$ has the component along $B$. 
In all three of the above modes, the magnetic fields $\cal B$ are always orthogonal to $\vp$. 
On the other hand, only in the $v^\mu_{(1)}$ mode, the electric field $\cal E$ is always orthogonal to $\vp$. 
In the $v^\mu_{(0)}$ and $v^\mu_{(2)}$ modes, one finds 
$\vp\cdot { \overrightarrow{\cal E}}_{(0)} \propto p^2 \vp_\perp^2 $ and 
$\vp\cdot { \overrightarrow{\cal E}}_{(2)} \propto p^2 p^3$, 
so that they are orthogonal only when $\vp\parallel \vB$ and $\vp\perp \vB$, respectively.\footnote{
In Ref.~\cite{Hattori:2012je}, one finds that the electric field in the $v^\mu_{(2)}$ mode is orthogonal to $ \vp $ 
when $\vp\parallel \vB$ as well as $\vp\perp \vB$ at $T=0$. 
This is because, when $\vp\parallel \vB$, the on-shell condition becomes $ p^2 = p_\parallel^2 = 0 $ according to 
the boost invariance along the external $ B $. 
However, at finite temperature, $ p_\parallel^2 \not= 0 $ even when $\vp\parallel \vB$ 
due to the absence of the Lorentz symmetry. 
Indeed, as we will see in the next subsections, $\cal E$ and $\vp$ are parallel when $\vp \parallel \vB$.} 
The physical meaning of $v^\mu$ can be discussed in the same way as in Ref.~\cite{Hattori:2012je}. 
$v^\mu$ induces $\cal E$ that is parallel to $\vp$, while does not induce any $\cal B$.

%%%%%%%%%%%%%%%%%%%%%%%%%%%%%%%%%%%%%%%%%%%
\subsection{Special momenta}
\label{ssc:special-momentum}

To get a feeling on the physical meaning of Eq.~(\ref{eq:Dmunu-general}), 
let us consider the two special momentum configurations below.

%%%%%%%%%%%%%%%%%%%%%
\subsubsection{$\vp\parallel \vB$ case}

\begin{figure}[t] 
\begin{center}
\includegraphics[width=0.4\textwidth]{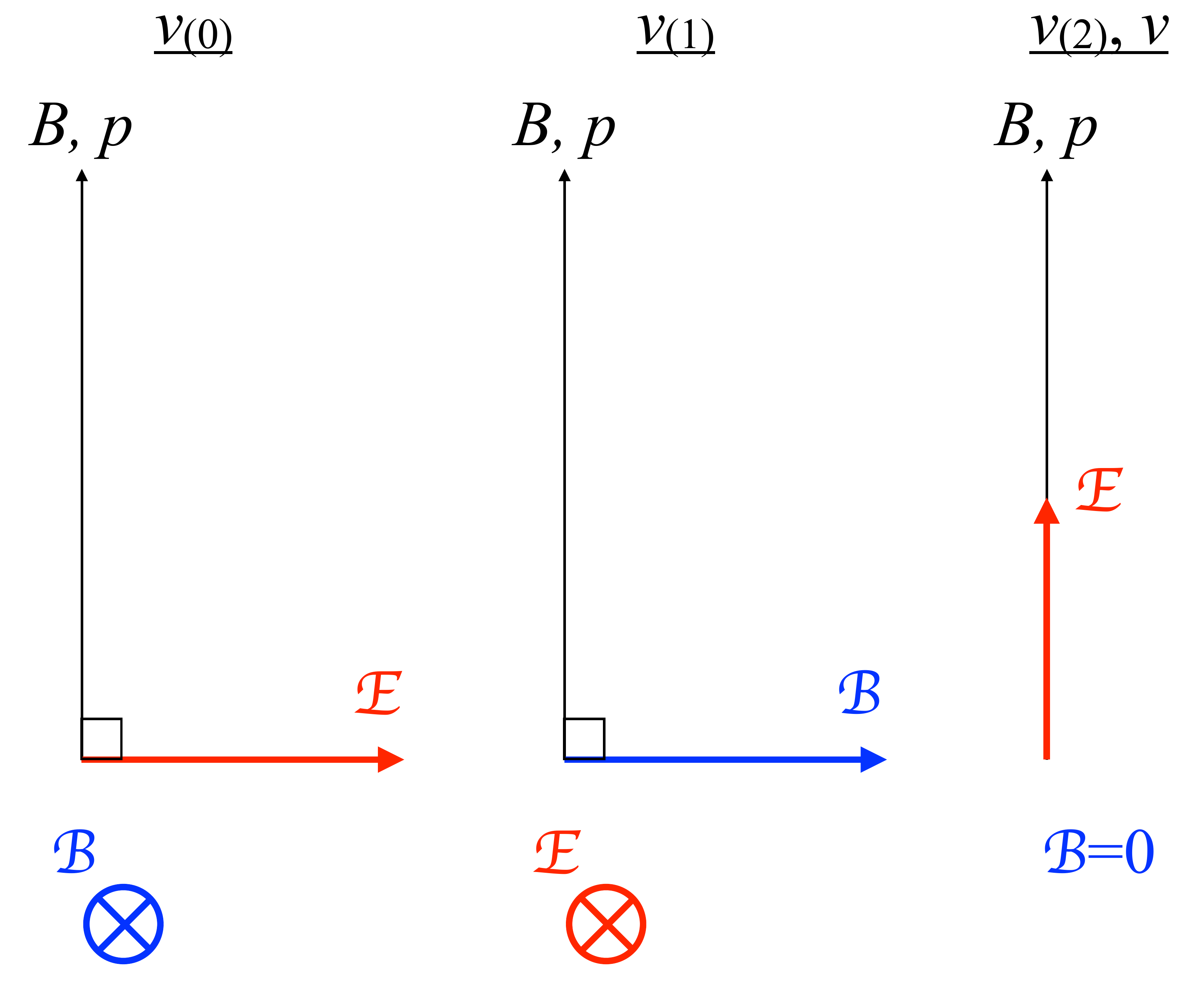} 
\caption{The directions of $\cal E$ and $\cal B$ generated by the four polarization vectors, in the case of $\vp\parallel \vB$.
} 
\label{fig:polarization-parallelp} 
\end{center} 
%\end{figure} 
%%%%%%%%%%%%%%%%%%%%%%%%%%%%%%%%%%%%%%%%%%%%
%\begin{figure}[t] 
\begin{center}
\includegraphics[width=0.4\textwidth]{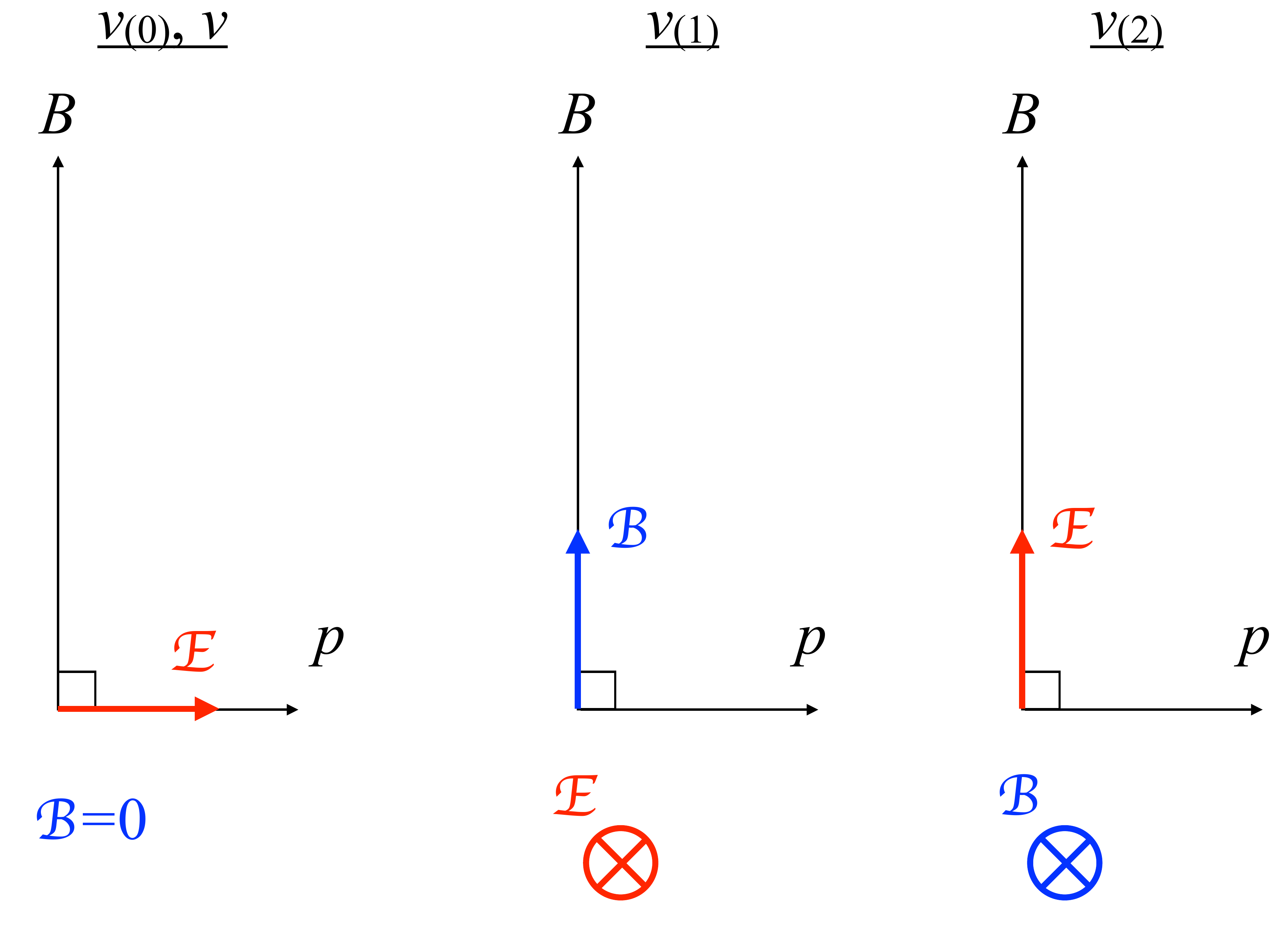} 
\caption{The directions of $\cal E$ and $\cal B$ generated by the four polarization vectors, in the case of $\vp\perp \vB$.
} 
\label{fig:polarization-perpp} 
\end{center} 
\end{figure}

When $\vp$ is parallel to the magnetic field, 
Eq.~(\ref{eq:Delta-general}) reduces to $\Delta= (p^2-\varPi_T)(p^2-\varPi_L  -\varPi_\parallel)$. 
Therefore, Eq.~(\ref{eq:Dmunu-general}) becomes 
\begin{align} 
\label{eq:D-pBparallel}
\begin{split}
D^R_{\mu\nu}(p)
&= -\Biggl[\frac{v^{(0)}_\mu v^{(0)}_\nu}{p^2-\varPi_T} 
 +\frac{v^{(1)}_\mu v^{(1)}_\nu}{p^2-\varPi_T-\varPi_\perp} \\
&~~~+ \frac{v^{(2)}_\mu v^{(2)}_\nu}{p^2-\varPi_L  -\varPi_\parallel}
\Biggr],
\end{split} 
\end{align} 
where we have used $v_{\mu}v_{\nu}=v^{(2)}_\mu v^{(2)}_\nu$ and $P^T_{\mu\nu}=v^{(0)}_\mu v^{(0)}_\nu+v^{(1)}_\mu v^{(1)}_\nu$. 
The correspondences between the polarization vectors 
and the directions of $\cal E$ and $\cal B$ are drawn in Fig.~\ref{fig:polarization-parallelp}. 
We see that $v^{(2)}_\mu$ coincides with the polarization vector $v^\mu$ longitudinal to the momentum $\vp$, 
while $v^{(0)}_\mu$ and $v^{(1)}_\mu$ are responsible for the two transverse polarizations.
The fact that $\varPi_T$ and $\varPi_\perp$ ($\varPi_L$ and $\varPi_\parallel$) are simply added in the denominator of the $v^{(1)}_\mu$ ($v^{(2)}_\mu$) term in the expression above can be understood from this fact.

In this configuration, the two transverse excitations specified by $v^{(0)}_\mu$ and $v^{(1)}_\mu$ are expected to be degenerated due to the rotational symmetry around the $B$ axis. 
It implies the limiting behavior $\varPi_\perp=0$ at $\vp_\perp=\vzero$.

%%%%%%%%%%%%%%%%%%%%%%%%%%%%%%%%%%%%%%%%%%%

%%%%%%%%%%%%%%%%%%%%%%%%%%%%%%%%%%%%%%%%%%%

\subsubsection{$\vp\perp \vB$ case}

When $\vp$ is perpendicular to $\vB$, Eq.~(\ref{eq:Delta-general}) reduces to $\Delta= (p^2-\varPi_L)(p^2-\varPi_T  -\varPi_\parallel)$. 
Therefore, Eq.~(\ref{eq:Dmunu-general}) becomes
\begin{align} 
\label{eq:D-pBperp}
\begin{split}
D^R_{\mu\nu}(p)
&= -\Biggl[\frac{v^{(0)}_\mu v^{(0)}_\nu}{p^2-\varPi_L}
+\frac{v^{(1)}_\mu v^{(1)}_\nu}{p^2-\varPi_T-\varPi_\perp} \\
&~~~ +\frac{v^{(2)}_\mu v^{(2)}_\nu}{p^2-\varPi_T  -\varPi_\parallel}
\Biggr] ,
\end{split}
\end{align} 
where we have used $v_\mu v_\nu=v^{(0)}_\mu v^{(0)}_\nu$ and $P^T_{\mu\nu}=v^{(1)}_\mu v^{(1)}_\nu+v^{(2)}_\mu v^{(2)}_\nu$.
The directions of the color-electromagnetic fields are summarized in Fig.~\ref{fig:polarization-perpp}.
From this figure, it is clear that the excitation specified by $v^{(0)}_\mu$ 
coincides with the longitudinal polarization, 
while those specified by $v^{(1)}_\mu$ and $v^{(2)}_\mu$ correspond to the two transverse excitations. 
Along with these correspondences, one can understand the fact in Eq.~(\ref{eq:D-pBperp}) 
that $\varPi_L$ appears only in the $v^{(0)}_\mu$ channel, 
while $\varPi_T$ appears in the $v^{(1)}_\mu$ and $v^{(2)}_\mu$ channels 
together with $\varPi_\parallel$ and $\varPi_\perp$, respectively.

\section{One-loop approximation}
\label{sec:one-loop}

\begin{figure}[t] 
\begin{center}
\includegraphics[width=0.2\textwidth]{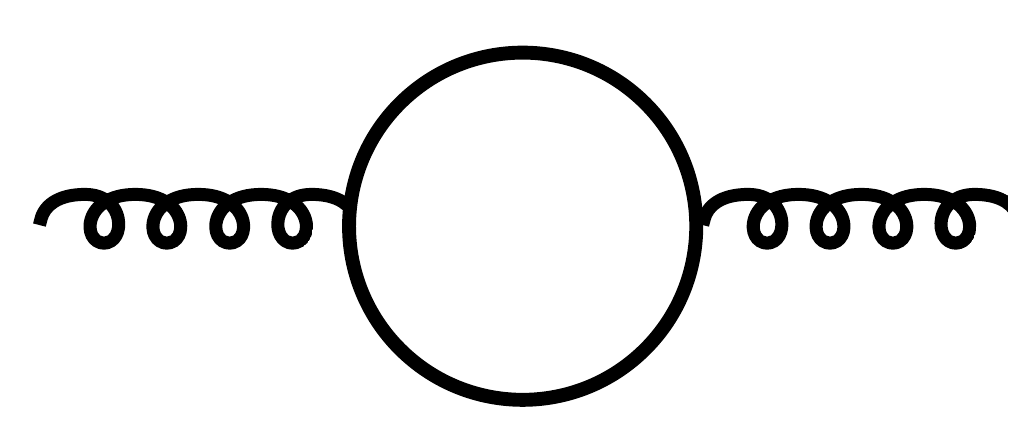} 
\caption{The one-loop diagram for quark contribution to the gluon self-energy.
The solid (curly) line represents a quark (gluon) propagator.
The quark is confined in the LLL.
} 
\label{fig:gluon-selfenergy-q} 
\end{center} 
\end{figure}

\begin{figure}[t] 
\begin{center}
\includegraphics[width=0.49\textwidth]{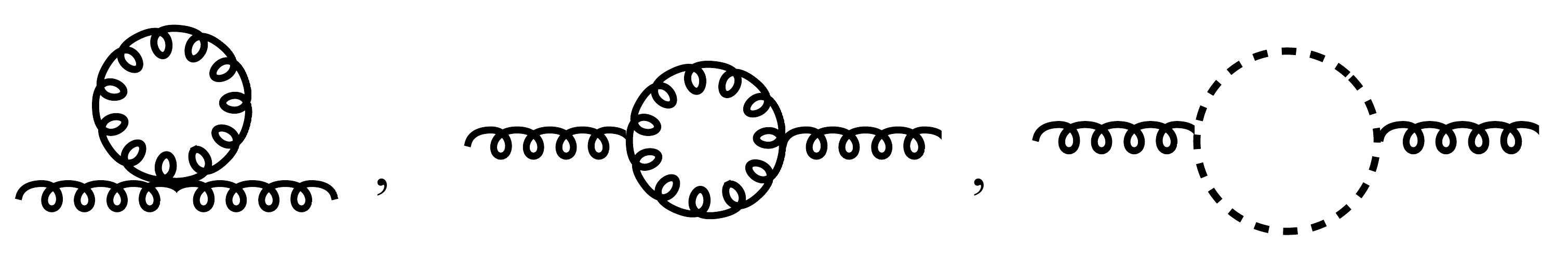} 
\caption{The one-loop diagram for gluon and ghost contributions to the gluon self-energy.
The dotted line represents the ghost propagator.
} 
\label{fig:gluon-selfenergy-g} 
\end{center} 
\end{figure}

In this section, we consider a specific self-energy 
which is given by the one-loop diagrams drawn in Figs.~\ref{fig:gluon-selfenergy-q} and~\ref{fig:gluon-selfenergy-g}.
From now on, we consider the strong magnetic field case, $\sqrt{eB}\gg T$, where $e$ is the electromagnetic coupling constant, so that the LLL approximation is valid.
In this approximation, the quark-loop contribution is proportional to the tensor $P^{\mu\nu}_\parallel$~\cite{Fukushima:2011nu}.
On the other hand, the gluon and ghost loops are not directly affected by the magnetic field, 
so that their tensor structures are the same as those at $B=0$:
They have only the components for $P^{\mu\nu}_T$ and $P^{\mu\nu}_L$. 
In the LLL, the fluctuation transverse to the magnetic field is absent, 
so that one can set $\varPi_\perp=0$ in Eq.~(\ref{eq:Dmunu-general}). 
We also consider the massless quarks for simplicity.\footnote{
In the LLL quark-loop contribution to the polarization tensor, the leading mass/temperature correction to the exact result $\sim g^2 eB $ from the massless Schwinger model 
is suppressed by $(m/T)^2  $ for a small quark mass 
$ m \ll T $ (see, e.g., Ref.~\cite{Fukushima:2015wck} and references therein). 
Therefore, the parametric dependence of the correction term is $\sim g^2 eBm^2/T^2$, and the temperature dependence is $T^{-2}$ instead of $T^2$, unlike in the usual four-dimensional case. }
In this case, we have $\varPi_\parallel(p)=M^2\equiv g^2\sum_f |\Bf|/(2\pi)^2$~\cite{Fukushima:2011nu}, 
where $f$ is the index for the quark flavor, 
and $\Bf\equiv eq_f B$ with $q_f$ being the electric charge of the quark carrying the flavor $f$.
Because of $\sqrt{eB}\gg T$, one finds $\varPi_\parallel \gg \varPi_{T,L} \sim g^2T^2$.

These inequalities naturally introduce a hierarchy of the energy scales: 
$gT\ll M\sim g\sqrt{eB} \ll T \ll \sqrt{eB}$. 
We will discuss the gluon spectrum in each scale, $p\sim gT$, $g\sqrt{eB}$, and $p\gg g\sqrt{eB}$. 
We will show a novel excitation in the first energy region. 
While the results in the last two cases are already known, 
we will also briefly discuss these cases in order to present 
a complete and self-contained discussion in the whole energy region. 

Note that the coupling constants in $ gT $ and $M  $ 
come from the gluon/ghost and quark loops, respectively, 
so that natural scales for the evaluation of the running QCD coupling constant may be different: 
A natural scale of the quark dynamics could be either of 
the inverse of the cyclotron radius $ \sim \sqrt{eB}  $ appearing in the transverse dynamics or $T$ appearing in the longitudinal dynamics, 
while that of the gluon/ghost dynamics is not subject to the size of the magnetic field and therefore $T$ is the natural scale.
To decide the appropriate energy scale in the running coupling in the quark dynamics, one needs to perform the higher-order calculation and try to minimize its contribution by choosing the energy scale. 
Nevertheless, in this paper, in each region specified above, 
we will only have the coupling constant either from $ gT $ or $M  $, 
which, therefore, can be evaluated with one scale.

%%%%%%%%%%%%%%%%%%%%%

%%%%%%%%%%%%%%%%%%%%%
\subsection{$p\gg M$ region}

When $p$ is so large that all of the self-energy corrections are negligible, 
the propagator reduces to the one in the free limit (\ref{eq:D-free}).
Equation~(\ref{eq:D-free}) is rewritten as 
\begin{align}
D^0_{\mu\nu}(p)
&= -\frac{v^\mu_{(0)} v^\nu_{(0)}+v^\mu_{(1)} v^\nu_{(1)}+v^\mu_{(2)} v^\nu_{(2)}}{p^2}.
\end{align}
As mentioned before, each $v^\mu_{(i)}$ is the polarization vector of 
the real-gluon field $A_\mu$~\cite{Hattori:2012je}, 
and its dispersion relation is given by the pole position of the corresponding term in $D_{\mu\nu}$. 

In the current case, the dispersion relations of the three modes ($i=0,1,2$) 
are all degenerated with $p^2=0$. 
The mode for $i=0$ was shown to be unphysical~\cite{Hattori:2012je} 
when the dispersion relation is light-like $p^2=0$. 
Thus, the number of the physical modes in this case is two, 
as it should be because a real gluon is allowed to have 
only the two transverse modes in the Lorentz-symmetric case.

%%%%%%%%%%%%%%%%%%%%%
\subsection{$p\sim M$ region}

When $p\sim M$, we have $p^2 \sim \varPi_\parallel \sim (g\sqrt{eB})^2 \gg \varPi_{T, L} \sim (gT)^2$.
Then, Eq.~(\ref{eq:Delta-general}) is approximated as $\Delta\simeq p^2\left(p^2 -\varPi_\parallel \right)$, while $D_\perp(p) \simeq 0$.
The resultant gluon propagator reads 
\begin{align} 
\begin{split}
D^R_{\mu\nu}(p)
&\simeq -\frac{v^\mu_{(0)} v^\nu_{(0)}+v^\mu_{(1)} v^\nu_{(1)}}{p^2} 
-\frac{v^\mu_{(2)} v^\nu_{(2)}}{p^2 - M^2}, 
\end{split}
\end{align}
where we have used $P^{\mu\nu}_T+P^{\mu\nu}_L=P^{\mu\nu}_0$.
This expression agrees with the known result~\cite{Li:2016bbh}.
We see that the two modes, which correspond to $v^\mu_{(0)}$ and $v^\mu_{(1)}$, 
are not affected by the interactions and have the unscreened dispersion relation $p^2=0$, 
while the other mode $v^\mu_{(2)}$ is screened and has the modified dispersion relation $p^2=M^2$.

%%%%%%%%%%%%%%%%%%%%%%%%%%%%%%%%%%%%%%%%%%%
\subsection{$p\sim gT$ region (HTL scale)}
\label{ssc:soft}

In this subsection, we focus on the soft-energy region, $p\sim gT$.
In this case, we can use the HTL approximation~\cite{Frenkel:1989br, Braaten:1990az} for $\varPi_T$ and $\varPi_L$:
\begin{align}
\varPi_T(p)
&= \frac{3}{2}\omega^2_p 
\left[y^2+(1-y^2)Q(y) \right] ,\\
\varPi_l(p)
&= 3\omega^2_p\left[-1+Q(y)\right] ,
\end{align}
where $y\equiv p^0/|\vp|$ and 
\begin{align}
Q(y)&\equiv \frac{y}{2}\left[\ln\left|\frac{1+y}{1-y}\right|-i\pi\theta(1-y^2)\right] .
\end{align}
We have defined $\varPi_l\equiv \vp^2 \varPi_L(p) /p^2$ for future convenience. 
%The gluon plasma frequency is given by $\omega_p \equiv gT \sqrt{\Nc}/3$ (with $\Nc$ being the number of colors), 
%which is different from the one at $B=0$ because the quark contribution is not included in the present case.
Notice that the quark-loop contribution is not included in the above HTL results. 
Consequently, the plasma frequency, which is now only from the gluon/ghost-loop contribution, 
is given by $\omega_p \equiv gT \sqrt{\Nc}/3$ (with $\Nc$ being the number of colors), 
which differs from the one at $B=0$ by overall factors. 

Since we are looking at the scale $p\sim gT$, 
the orders of the self-energies are $\varPi_{T, L} \sim (gT)^2$ and $\varPi_\parallel\sim (g\sqrt{eB})^2$.
Thus, Eqs.~(\ref{eq:Delta-general}) and (\ref{eq:Dperp-general}) are approximated as 
$\Delta\simeq -\varPi_\parallel p^2  \Delta_{\text{mix}}/p^2_\parallel$ and 
\begin{align}
D_\perp(p) &\simeq 
\frac{1}{p^2-\varPi_T} 
\varPi_\parallel(\varPi_L-\varPi_T)(1-a)\frac{(p^0)^2}{p^2_\parallel} ,
\end{align}
where 
\begin{align}
\Delta_{\text{mix}}\equiv p^2_\parallel-\varPi_T a -\varPi_l y^2(1-a).
\end{align}
By using these expressions, Eq.~(\ref{eq:Dmunu-general}) results in 
\begin{align} 
\label{eq:D-soft}
\begin{split}
D^R_{\mu\nu}(p)
&\simeq \frac{1}{\Delta}
\Bigl[\varPi_\parallel P^0_{\mu\nu}(p) 
-\varPi_\parallel P^\parallel_{\mu\nu}(p) 
-D_\perp(p)P^\perp_{\mu\nu}(p)
\Bigr]\\
& = - \frac{p^2_\parallel}{p^2} \frac{ v^{(0)}_\mu v^{(0)}_\nu }{ \Delta_{\text{mix}}} 
 -\frac{v^{(1)}_\mu v^{(1)}_\nu }{p^2-\varPi_T}.
\end{split}
\end{align}
In the above, the dependence on $\varPi_\parallel$ goes away 
because of the cancellation between the numerator and the denominator. 
We note that the mode with $v^{(2)}_\mu$ does not appear here 
because this mode is, as we have seen in the previous subsection, screened by the screening mass $M$, and thus it has much larger energy than the one we are considering ($\sim gT$).

The first term of Eq.~(\ref{eq:D-soft}) reduces to the purely transverse (longitudinal) component 
when $\vp\parallel \vB$ ($\vp\perp \vB$):
\begin{subequations}
\begin{align}
\label{eq:Deltamix-T}
&\Delta_{\text{mix}}= p^2-\varPi_T 
~~~~~~~~~(\vp\parallel \vB),\\
\label{eq:Deltamix-L}
&\Delta_{\text{mix}}=y^2\left[ \vp^2 -\varPi_l \right]
~~~(\vp\perp \vB).
\end{align}
\end{subequations}
They agree with the two denominators appearing in the HTL propagator~\cite{Frenkel:1989br, Braaten:1990az} 
in the absence of the magnetic field, 
which can be obtained by setting $\varPi_\parallel=\varPi_\perp=0$ in Eq.~(\ref{eq:Dmunu-general}):
\begin{align} 
\label{eq:D-noB-HTL}
\begin{split}
D^R_{\mu\nu}(p)
&= -\frac{P^T_{\mu\nu}(p) }{p^2-\varPi_T}
 -\frac{\vp^2}{p^2}\frac{P^L_{\mu\nu}(p)}{\vp^2-\varPi_l}.
\end{split}
\end{align}
For general values of $a$, $\Delta_{\text{mix}}$ is a linear combination of $\varPi_T$ and $\varPi_l$.
In other words, the transverse and the longitudinal components are ``mixed'', 
which does not happen in the $B=0$ case.

On the other hand, the denominator of the second term in Eq.~(\ref{eq:D-soft}) 
is the same as that of the transverse term of Eq.~(\ref{eq:D-noB-HTL}), so this term does not have novel properties.

%%%%%%%%%%%%%%%%
\subsubsection{collective excitations}

In the time-like region $p^2>0$, $\varPi_T$ and $\varPi_l$ do not have imaginary parts. 
Therefore, Eq.~(\ref{eq:D-soft}) can have poles on the real axis in the $p^0$ plane. 
If they exist, they correspond to collective excitations of which the energies are given by the pole positions.

Let us begin with the two special cases, $\vp\parallel \vB$ and $\vp\perp \vB$, 
which correspond to $a=1  $ and $a=0  $, respectively. 
In these cases, $\Delta_{\text{mix}}$ reduces to the well-known HTL results at $B=0$ 
up to the absence of the quark-loop contribution in the present case.

\begin{figure}[t] 
\begin{center}
\includegraphics[width=0.5\textwidth]{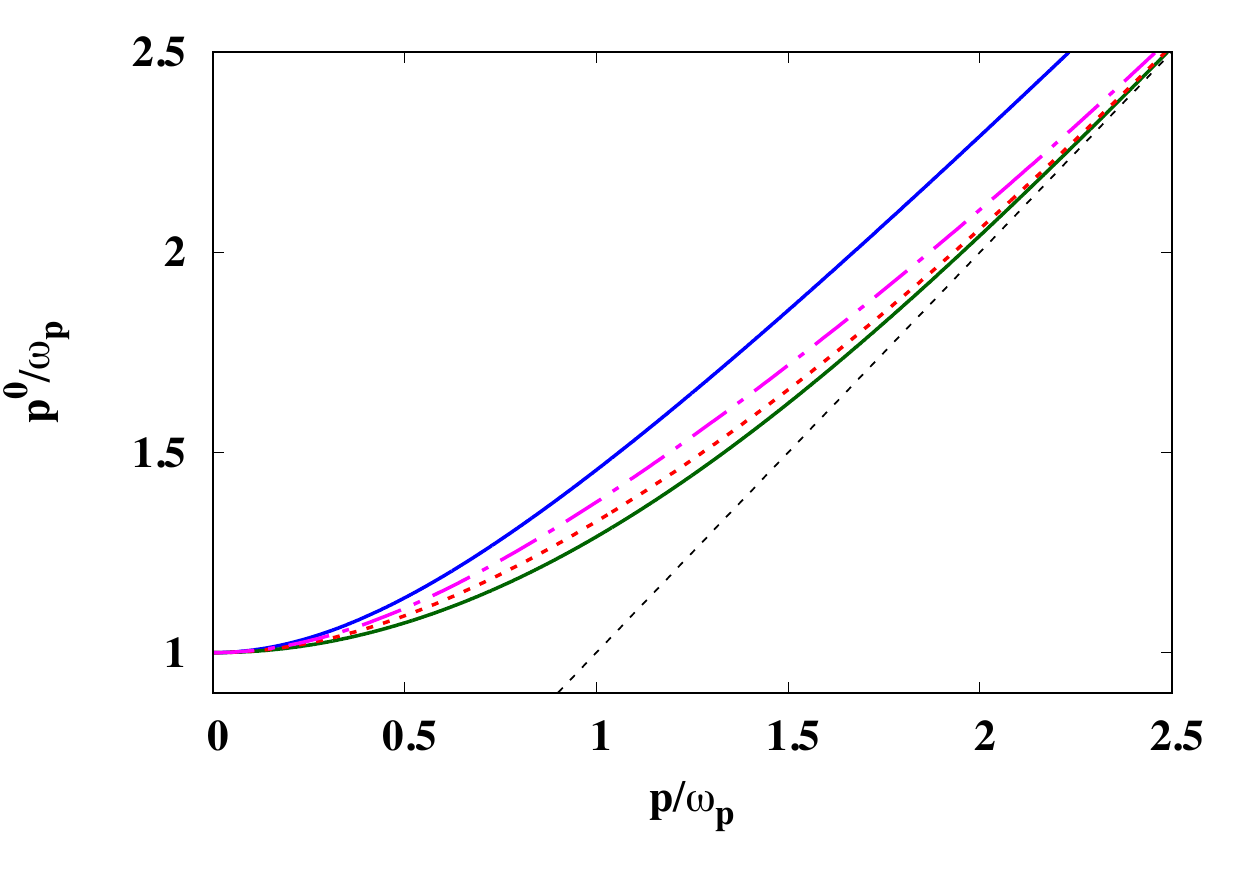} 
\caption{The pole position obtained from $\Delta_{\text{mix}}=0$ with $a=0.3$ (red, dotted line) and $a=0.6$ (magenta, chain line), 
as a function of the momentum.
The poles in the transverse (blue, solid line) and the longitudinal (green, solid line) components 
at $B=0$ (Eq.~(\ref{eq:D-noB-HTL})), 
which correspond to $a=1$ and $a=0$ in $\Delta_{\text{mix}}=0$, respectively, are also plotted. 
The light cone (black, dotted line) is also plotted. 
The unit of the energy and of the momentum is $\omega_p$. 
} 
\label{fig:dispersion} 
\end{center} 
\end{figure}

In the former case at $ a=1 $, 
the root of Eq.~(\ref{eq:Deltamix-T}) is plotted\footnote{Note that we plotted only the pole with the positive energy.
Another pole with the same magnitude but with the opposite sign exists in the negative energy region, as can be seen from the symmetry of $\varPi_T$ under the transformation $p^0\rightarrow -p^0$.}
in Fig.~\ref{fig:dispersion} as a function of $|\vp|$.
This also gives the pole of the second term in Eq.~(\ref{eq:D-soft}), 
because the dispersion relations of the two transverse modes should be degenerated 
according to the rotational symmetry with respect to the direction of the magnetic field. 
Its asymptotic forms can be analytically obtained as~\cite{Frenkel:1989br, Braaten:1990az}
\begin{subequations}
\begin{align}
(p^0)^2&\simeq \omega^2_p+\frac{6}{5}\vp^2
~~(|\vp| \ll gT),\\
\label{eq:pole-T-asymptotic}
(p^0)^2&\simeq \vp^2+\frac{3}{2}\omega^2_p
~~(|\vp| \gg gT) ,
\end{align}
\end{subequations}
where we have used $Q(y)\simeq 1+1/(3y^2)+1/(5y^4)$ for $y \gg 1$, 
and $Q(y)\simeq \ln(2/\epsilon)/2$ for $y\simeq 1+\epsilon$ with $\epsilon \ll 1$.

In the latter case at $a=0  $, $\Delta_{\text{mix}}$ has only the longitudinal component.
The root of Eq.~(\ref{eq:Deltamix-L}) is plotted in Fig.~\ref{fig:dispersion} as a function of $|\vp|$.
This collective excitation is known as the plasmon.
Its asymptotic forms are obtained as~\cite{Frenkel:1989br, Braaten:1990az}
\begin{subequations}
\begin{align}
& (p^0)^2\simeq \omega^2_p+\frac{3}{5}\vp^2 
\hspace{2.7cm} (|\vp| \ll gT),\\
& p^0 \simeq 
|\vp|\left[1+2\exp\left(-2-\frac{2\vp^2}{3\omega^2_p}\right)\right]
~~(|\vp| \gg gT) .
\end{align}
\end{subequations}
We note that the dispersion curve quickly approaches the light cone at the large $|\vp|$ 
due to the exponential factor. 
This contrasts to the asymptotic behavior of  the transverse component shown in Eq.~(\ref{eq:pole-T-asymptotic}).

Now, we move to general values of $a$. In Fig.~\ref{fig:dispersion}, 
the solutions of $ \Delta_{\text{mix}}=0$ are plotted\footnote{Here we do not consider the pole at $p^2=0$, which originates from the prefactor of the first term in Eq.~(\ref{eq:D-soft}), because such pole with the polarization vector $v^{(0)}_\mu$ is unphysical~\cite{Hattori:2012je}.} for $a=0.3$ and $0.6$ as functions of $|\vp|$.
The pole positions appear inbetween those of the transverse channel, $p^2-\varPi_T=0$, 
and of the longitudinal channel, $\vp^2-\varPi_l=0$, at $B=0$. 
One can see that, as $a$ increases from zero, 
the pole position departs from that in the longitudinal channel, 
and approaches that in the transverse channel. 
This dispersion relation at the fractional value of $  a$ 
is different from both of the transverse and the longitudinal channels at $B=0$, 
and thus is a novel collective excitation appearing only in the presence of the strong magnetic field. 
We can obtain the asymptotic forms of the pole position as 
\begin{subequations}
\begin{align}
&(p^0)^2 \simeq \omega^2_p+\frac{3(a+1)}{5}\vp^2
~~(|\vp| \ll gT),\\
&p^0\simeq 
|\vp|\left[1+2\exp\left(-2-\frac{2\vp^2}{3\omega^2_p}+\frac{a}{1-a}\right)\right]
~~(|\vp| \gg gT).
\end{align} 
\end{subequations}
At the large $|\vp|$, the dispersion curve approaches the light cone exponentially, 
just like that of the plasmon at $B=0$.
We note that, when $a$ is very close to unity, 
the latter asymptotic expression becomes invalid as noticed from the blow up of the exponential, 
and one should use Eq.~(\ref{eq:pole-T-asymptotic}) instead.

\begin{figure}[t] 
\begin{center}
\includegraphics[width=0.5\textwidth]{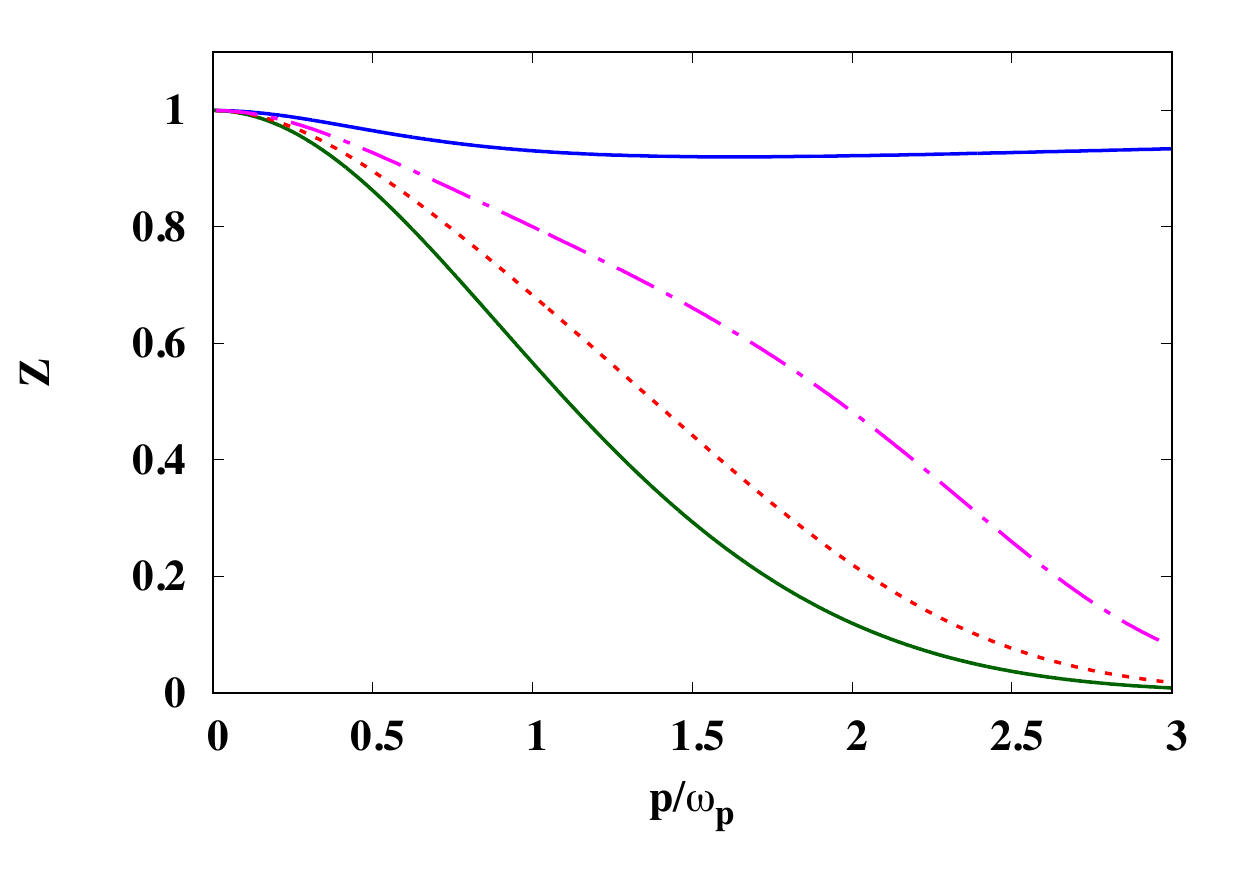} 
\caption{The residues $Z$ at $a=0.3$ (red, dotted line) and $a=0.6$ (magenta, chain line), as a function of the momentum. 
The residues at $a=1$ (blue, solid line) and $a=0$ (green, solid line), 
which correspond to the transverse and the longitudinal channels at $B=0$, respectively, 
are also plotted for comparison. The unit of the momentum is $\omega_p$.
} 
\label{fig:residue} 
\end{center} 
\end{figure}

Next, we discuss the residue of the novel collective excitation.
By expanding the first term in Eq.~(\ref{eq:D-soft}) around the pole [which we write as $p^0=\omega(|\vp|$)] as 
\begin{align} 
\begin{split}
\frac{1 }{ \Delta_{\text{mix}}} 
&\simeq \frac{Z(|\vp|)}{2\omega(|\vp|)}
\frac{1}{p^0-\omega(|\vp|)},
\end{split}
\end{align}
 we introduce the residue
\begin{align}
Z(|\vp|)&= 2\omega(|\vp|)
\left(\left. \frac{d \Delta_{\text{mix}}}{dp^0}
\right|_{p^0=\omega(|\vp|)}\right)^{-1} .
\end{align}
As in the discussion for the pole positions, we start with the two special cases.
In Fig.~\ref{fig:residue}, we show the residue in the $\vp\parallel \vB$ case 
where the excitation reduces to the transverse one at $B=0$. 
We see that the residue does not change much in the all momentum range.
Especially, in the both limits $|\vp|\rightarrow 0$ and $|\vp|\rightarrow \infty$, 
it is known that the residue approaches unity~\cite{Frenkel:1989br, Braaten:1990az}. 
On the other hand, the residue in the $\vp\perp \vB$ case, plotted in the same figure, 
corresponds to the plasmon excitation at $ B=0 $. 
In this case, it is also known that, whereas the residue approaches unity as $|\vp| \to 0$, 
it decreases exponentially at the large $|\vp|$~\cite{Frenkel:1989br, Braaten:1990az}. 
Therefore, contrary to the transverse excitations, 
the residue shows a rapid change as a function of $ |\vp| $.

We shall move to general values of $a$. 
In Fig.~\ref{fig:residue}, we show the residues as functions of $|\vp|$ for $a=0.3$ and $0.6$. 
An analytical expression of the residue at the small $|\vp|$ is obtained as 
\begin{eqnarray}
Z\simeq 1- \frac{ 3-2a}{5}  \left(\frac{|\vp|}{\omega_p} \right)^2
\, .
\end{eqnarray}
We find that, for the general $  a$, the residue goes to unity as $ |\vp| \to 0 $. 
The asymptotic form at the large $|\vp|$ can also be obtained as 
\begin{align}
Z&\simeq \frac{8\vp^2}{3(1-a)\omega^2_p}
\exp\left(-2-\frac{2\vp^2}{3\omega^2_p}
+\frac{a}{1-a}\right).
\end{align}
The exponential suppression at the large $|\vp|$ suggests that 
the excitation does not have physical significance when $|\vp|\gg gT$, 
where the interaction effect can be safely neglected. 
These analytic expressions confirm the behaviors both 
in the small and large $|\vp|$ regions in our numerical results. 

At the large $|\vp|$, we see that the behaviors at finite $  a$ 
are more similar to that of the plasmon rather than of the transverse excitation at $B=0$, 
suggesting that the new excitation is a collective excitation like the plasmon at $B=0$.

%%%%%%%%%%%%%%%%%%%%%%%%%%%%%%%%%%%%%%%%%%%

%%%%%%%%%%%%%%%%%%%%%%%%%%%%%%%%%%%%%%%%%%%

%%%%%%%%%%%%%%%%
\subsubsection{Debye screening}

The limiting behavior of the first term in Eq.~(\ref{eq:D-soft}) at $p^0=0$ 
describes the screening property of static color-electric field.
We start with the $\vp\parallel \vB$ case.
At $p^0=0$, we have $\varPi_T=0$, so $p^2_\parallel/[p^2\Delta_{\text{mix}}]=-1/\vp^2$.
It indicates the well-known fact that there is no static screening effect in the transverse channel.
By contrast, for the $\vp\perp \vB$ case, 
we have $p^2_\parallel/[p^2\Delta_{\text{mix}}]=-1/[\vp^2+3\omega^2_p]$ from $\varPi_l=-3\omega^2_p$. 
This expression indicates the screening effect in the longitudinal channel, 
of which the Debye mass is given by $\sqrt{3}\omega_p$~\cite{Frenkel:1989br, Braaten:1990az}.

Now, let us move to the case of general $a$.
The first term in Eq.~(\ref{eq:D-soft}) becomes
\begin{align}
\frac{p^2_\parallel}{p^2\Delta_{\text{mix}}}
&= -\frac{1}{\vp^2},
\end{align}
at $p^0 = 0$.
This expression suggests that there is no static screening effect 
except for the $\vp\perp \vB$ case ($a = 0)$. 
It is surprising that the static screening disappears 
once $a$ gets an infinitesimal deviation from zero.

%%%%%%%%%%%%%%%%
\subsubsection{Dynamical screening}

Although the static screening is absent, 
there can be a dynamical screening effect in the first term of Eq.~(\ref{eq:D-soft}).
We focus on the behavior at the small $p^0$.
Let us start with the $\vp\parallel \vB$ case. 
By using $\varPi_T\simeq -3i\pi y\omega^2_p/4$ for $p^0\ll |\vp|$, which comes from the Landau damping, we get $p^2_\parallel/[p^2\Delta_{\text{mix}}]\simeq -1/[\vp^2-3i\pi p^0\omega^2_p/(4|\vp|)]$.
Its absolute value becomes 
\begin{align}
\left|\frac{p^2_\parallel}{p^2\Delta_{\text{mix}}} \right|^2
\simeq  \frac{1}{|\vp^2|^2+[3\pi p^0\omega^2_p/(4|\vp|)]^2} .
\end{align}
This expression shows the presence of the screening effect at finite $p^0$, which is called the dynamical screening.
This effect is quite important in the calculation of the transport coefficients~\cite{Arnold:2000dr, Arnold:2003zc, Gagnon:2007qt, Gagnon:2006hi} and the quark damping rate~\cite{Blaizot:1996az, Blaizot:1996hd, Lebedev:1990kt, Lebedev:1990un, Pisarski:1993rf} at $B=0$.
There is a dynamical screening effect also in the longitudinal channel, but the static screening effect dominates when $p^0\ll |\vp|$, so we do not discuss the $\vp\perp \vB$ case.

For general values of $a$, the first term of Eq.~(\ref{eq:D-soft}) becomes
\begin{align}
\begin{split}
\frac{p^2_\parallel}{p^2\Delta_{\text{mix}}}
&\simeq -\frac{1}{\vp^2+\varPi_T},
\end{split}
\end{align}
for $p^0\ll |\vp|$.
Interestingly, this expression shows no dependence on $a$ 
and is the same as that in the $\vp\parallel \vB$ case, 
where the expression reduces to that in the transverse channel at $B=0$. 
Therefore, it is clear for the general $  a $ 
that there is the same dynamical screening effect as that of the transverse mode. 

%%%%%%%%%%%%%%%%%%%%%%%%%%%%%%%%%%%%%%%%%%%
\subsection{Brief Summary of whole momentum ranges}

\begin{figure}[t] 
\begin{center}
\includegraphics[width=0.5\textwidth]{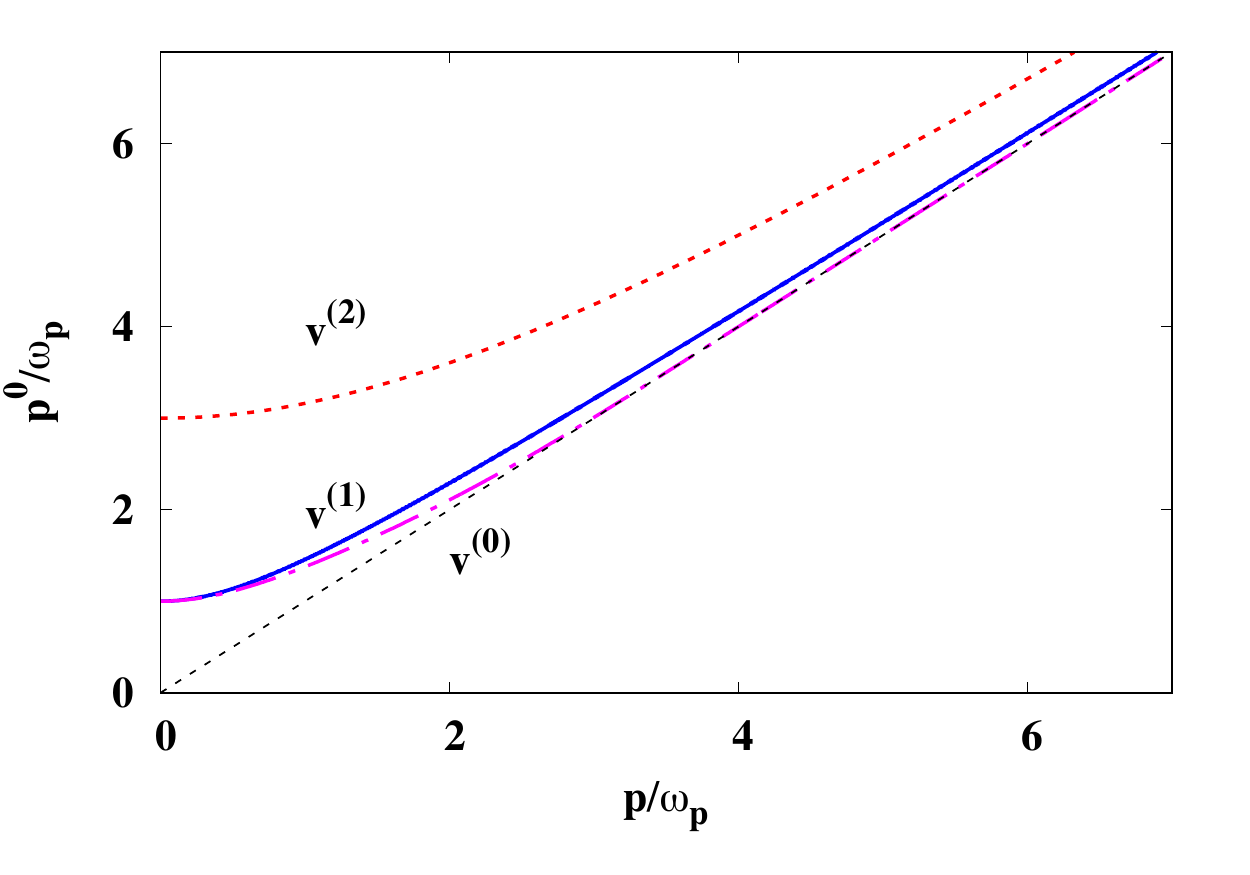}  
\caption{The dispersion relations of the three excitations.
The excitation with the polarization vectors $v^\mu_{(0)}$, $v^\mu_{(1)}$, and $v^\mu_{(2)}$ are represented by chain line colored in magenta, solid line colored in blue, and the dotted line colored in red, respectively.
The light cone (black, dotted line) is also plotted.
We set $M/\omega_p=3.0$ and $a=0.6$.
The unit for the energy and the momentum is $\omega_p$.
} 
\label{fig:dispersion-whole} 
\end{center} 
\end{figure} 

Here, we briefly summarize our results for the gluon spectrum obtained in this section. 
We show the dispersion relations of the excitations in Fig.~\ref{fig:dispersion-whole} 
which correspond to the polarization vectors $v^\mu_{(i)}$ ($i=0,1,2$), respectively. 

The excitation with $v^\mu_{(0)}$ has a pole at $p^0=\omega_p\sim gT$ at $|\vp|=0$, 
and the pole position increases as $|\vp|$ increases, and finally approaches the light cone when $|\vp|\gg gT$.
However, at the same time, the residue decreases exponentially at $|\vp|\gg gT$, 
so this excitation is physically important only when $|\vp|\lesssim gT$.
We note that, this picture does not hold when $a$ is very close to unity. 
In this case, $\varPi_l$ does not appear in $\Delta_{\text{mix}} $ as seen in Eq.~(\ref{eq:Deltamix-T}), 
and the pole position and the residue of this excitation become identical to 
those of the excitation with the polarization vector  $v^\mu_{(1)}$, of which the properties are discussed below.

Another excitation with $v^\mu_{(1)}$ also has a pole at $p^0=\omega_p\sim gT$ at $|\vp|=0$.
The pole position increases as the momentum increases, 
but does not approach the light cone in contrast to the excitation we have just discussed.
Its residue approaches unity at the large $|\vp|$, so it is still physical object also when $|\vp|\gg gT$.
Actually, the excitation with $v^\mu_{(1)}$ exists even at the larger momentum region, $|\vp|\gtrsim M$.

The other excitation with $v^\mu_{(2)}$ has the much larger mass compared with the other two:
At zero momentum, the pole is at $p^0=M\gg \omega_p$, and the pole position becomes $(p^0)^2=\vp^2+M^2$ at the finite momentum.
This excitation also exists at the momentum region $|\vp|\gg M$.

The two excitations with the polarization vectors $v^\mu_{(1)}$ and $v^\mu_{(2)}$ 
have finite residues at $|\vp|\gtrsim M$ where the interaction effect is negligible. 
The residue of the excitation with $v^\mu_{(0)}$ decreases exponentially with the increasing momentum, 
which indicates a clear contrast to the former two. 
This result suggests that the latter excitation is a purely collective excitation like the plasmon at $B=0$.

We also briefly comment on the screening properties.
As we have seen, the behavior of the residue for $v^\mu_{(0)}$ is similar to that in the longitudinal mode at $B=0$.
However, the screening properties of this component are the same as in the transverse mode at $B=0$:
Namely, the Debye screening is absent, while the dynamical screening effect persists. 
This is an intriguing fusion between the longitudinal and transverse modes 
caused by the magnetic field. 

As a final remark of this section, we comment on the gauge dependence of our result.
The gluon propagator itself is gauge-dependent quantity.
Actually, the general expression in the covariant gauge, Eq.~(\ref{eq:Dmunu-general}), is apparently different from the corresponding expression (\ref{eq:propagator-Coulomb}) in the Coulomb gauge.
Nevertheless, the physical quantities that can be calculated from the gluon propagator should be gauge-invariant, such as the pole position of the gluon collective excitation~\cite{Kobes:1990xf, Kobes:1990dc}.
Especially in our one-loop analysis, $\varPi_\parallel$ is independent of the gauge-fixing since only the quarks are involved inside the loop, and $\varPi_{T/L}$ is also gauge-independent because it is evaluated with the HTL approximation.
It proves that the properties of the collective excitations in this section, such as the dispersion relation and the strength, are gauge-independent.

%%%%%%%%%%%%%%%%%%%%%%%%%%%%%%%%%

\section{Summary
}
\label{sec:summary}

We obtained the gluon propagator at finite temperature 
and in a magnetic field, in which both of the Lorentz and rotational symmetries are broken,
 for the case that the gluon self-energy has the four independent transverse tensor components. 
%This is the most general expression, which is applicable at an arbitrary magnitude of the magnetic field. 

Then, by using the specific form of the one-loop gluon self-energy 
in the LLL approximation, which is valid in the strong magnetic field, 
we clarified the picture of the gluon excitations 
in the whole energy ranges, $p\gg M$, $p\sim M$, and $p\sim gT$. 
Especially in the $p\sim gT$ case, we found that there appear two collective excitations, 
and that the properties of one of them are significantly modified compared with those at $B=0$. 
We note that the gluon self-interaction is essentially important for the collective excitations at $p\sim gT$, 
so that these collective excitations do not appear in the relativistic 
quantum electrodynamics plasma 
because there are no counterparts for the diagrams in Fig.~\ref{fig:gluon-selfenergy-g}. 

We also discussed the Debye and dynamical screening effects, 
and found that the Debye screening is absent in contrast to the $B=0$ case, 
while the dynamical screening persists.
The expression of this dynamical screening will be useful 
when one computes the transport coefficients in the strong magnetic field, 
where the soft gluons appear as the exchanged particles. 

We note that, our calculation is based on the assumption of the strong magnetic field, $T \ll \sqrt{eB}$.
When this assumption is not realized, we can not rely on the LLL approximation so that $\varPi_\parallel$ is no longer equal to $M^2$, and $\varPi_\perp$ becomes finite and its magnitude would be the same order as the other components of $\varPi$.
Also, the inequality $\varPi_T$, $\varPi_L\ll \varPi_\parallel$ will not be realized any more.
The investigation in such case is left to future work. 

%%%%%%%%%%%%%%%%%%%%%%%%%%%%%%%%%%%%%%%%%%%%%%%%%
\section*{Acknowledgements}

The research of D.~S. is supported by Alexander von Humboldt Foundation. 
The research of K.~H.  is supported by China Postdoctoral Science Foundation under Grant No.~2016M590312. 

%%%%%%%%%%%%%%%%%%%%%%%%%%%%%%%%%%%%%%%%%%%%
\appendix
%\appendix*
\section{Gluon propagator in Coulomb gauge}
\label{app:Coulomb}

The bare propagator and its inverse in the Coulomb gauge are, respectively, given by  
\begin{align} 
D^0_{\mu\nu}(p)
&= -\frac{P^T_{\mu\nu}(p)}{p^2}
-\frac{n_\mu n_\nu}{\vp^2}
+\alpha\frac{p_\mu p_\nu}{(\vp^2)^2},
\\ 
[D^0]^{-1}_{\mu\nu}(p)
&= -p^2 P^0_{\mu\nu}(p)
+\frac{1}{\alpha} g_{\mu i} g_{\nu j} p^i p^j.
\end{align}
As for the tensor structure of the self-energy, 
we assume the %most general 
form in Eq.~(\ref{eq:selfenergy}). 

We consider the behavior in the limit $\alpha\rightarrow 0$.
By evaluating the terms of order $ \alpha^{-1} $ and $\alpha^0  $ 
in the definition of the inverse matrix, $D^{\mu\nu}_R [(D^0)^{-1}+\varPi]_{\nu\alpha}=g^\mu_\alpha$, we get
\begin{align}
\label{eq:Coulomb-eq1}
0&= p^i D^{\mu i}_R,\\
\label{eq:Coulomb-eq2}
g^\mu_\alpha
&= D^{\mu }_{g i} g_{\alpha j} p^i p^j
+ D^{\mu \nu}_R \left[-p^2 P^0_{\nu\alpha}
+\sum_{i=T, L, \parallel, \perp}\varPi_i(p) P^{\nu\alpha}_i \right],
\end{align}
where we have decomposed $D^{\mu\nu}_R$ as $D^{\mu\nu}_R+\alpha D^{\mu\nu}_g+{\cal O}(\alpha^2)$.
Generally, $D^{\mu\nu}_R$ has the tensor structure
\begin{align}
D^{\mu \nu}_R
&= \sum_{i=T, L, \parallel, \perp}D_i P^{\mu\nu}_i
+D_p\frac{p^\mu p^\nu}{p^2}+D_n n^\mu n^\nu +D_b b^\mu b^\nu ,
\end{align}
according to the argument in Sec.~\ref{ssc:gluon-general}. 
$D^{\mu\nu}_g$ also has the same tensor structure, 
and the coefficients in this order are referred to as $D^g_{i}$. 
Equation~(\ref{eq:Coulomb-eq1}) gives a constraint on these coefficients, resulting in 
\begin{align}
D^{\mu \nu}_R
&= \sum_{i=T, \perp}D_i P^{\mu\nu}_i
+D_n n^\mu n^\nu 
+D B^{\mu\nu} .
\label{eq2}
\end{align}
We have defined 
$B^{\mu\nu}\equiv P^{\mu\nu}_L-y^2 p^\mu p^\nu/p^2-(p^0)^2b^\mu b^\nu/(p^3)^2+p^2_\parallel P^{\mu\nu}_\parallel/(p^3)^2$ 
such that $ p^i B^{\mu i}=0 $.

By solving Eq.~(\ref{eq:Coulomb-eq2}), the gluon propagator is found to be 
\begin{align}
\label{eq:propagator-Coulomb}
\begin{split}
D^{\mu \nu}_R
&= -\frac{1}{\Delta}
\Biggl[ \left(p^2-\varPi_L-a\frac{p^2}{p^2_\parallel}\varPi_\parallel \right) P^{\mu\nu}_T 
-a\frac{p^2}{p^2_\parallel}\varPi_\parallel B^{\mu\nu} \\
&~~~+\frac{p^2}{\vp^2}n^\mu n^\nu\left(p^2-\varPi_T-\frac{(p^0)^2}{p^2_\parallel}\varPi_\parallel\right)
+D^c_\perp P^{\mu\nu}_\perp 
\Biggr],
\end{split}
\end{align}
where $\Delta$ is the same as that in the covariant gauge (\ref{eq:Delta-general}), and 
\begin{align}
\begin{split}
D^c_\perp &=-\left(p^2-\varPi_L-a\frac{p^2}{p^2_\parallel}\varPi_\parallel\right)
+\frac{\Delta}{p^2-\varPi_T-\varPi_\perp}.
\end{split}
\end{align}
The coefficients of $D^{\mu\nu}_g$ satisfy 
\begin{align}
D^g_L= \frac{(p^3)^2}{p^2_\parallel} D^g_\parallel
&= -\frac{(p^3)^2}{(p^0)^2} D^g_b ,\\
(p^0)^2 D^g_L+\vp^2 D^g_p&= \frac{p^2}{\vp^2} .
\end{align}
This result can be confirmed by using the properties of the four projection tensors and $B^{\mu\nu}$ summarized in Appendix~\ref{app:projection} as well as the definitions of the projection tensors (\ref{eq:PT-def})--(\ref{eq:Pperp-def}).

Before ending this Appendix, we examine the form of the resummed propagator near the mass-shell, which will be useful for the computation of the transport coefficients.
The on-shell condition is given by the denominator of the propagator, so in the free limit it reads $p^2=0$.
The interaction effect modifies the dispersion relation. 
However, the modification is of the order of $\varPi_i$, and thus $ p^2 $ is still small. 
For this reason, the coefficients of $n^\mu n^\nu$ and $B^{\mu\nu}$ in Eq.~(\ref{eq:propagator-Coulomb}) 
are negligible compared with the other tensors:
\begin{align}
\begin{split}
D^{\mu \nu}_R
&\simeq -\frac{p^2-\varPi_L}{\Delta}
\left(P^{\mu\nu}_T-P^{\mu\nu}_\perp \right)
-\frac{P^{\mu\nu}_\perp}{p^2-\varPi_T-\varPi_\perp}.
\end{split}
\label{eq11}
\end{align}
We have applied the same approximation to the coefficient of the term being proportional to $ P_T^{\mu\nu} $, 
and the denominator can be also approximated for $ p^2 \sim 0 $ as 
\begin{align}
\begin{split}
\Delta&\simeq (p^2-\varPi_T)(p^2-\varPi_L) 
 -\varPi_\parallel\left[ p^2-\varPi_T \frac{a}{1-a} \frac{p^2}{\vp^2}-\varPi_L\right] \\
&= (p^2)^2 
-p^2\left[\varPi_T+\varPi_L+\varPi_\parallel  -\varPi_\parallel \varPi_T \frac{a}{1-a} \frac{1}{\vp^2}     \right] \\
&~~~+\left(\varPi_T+\varPi_\parallel\right) \varPi_L.
\end{split}
\end{align}
Neglecting the term that is proportional to $p^2\varPi_\parallel \varPi_T$, 
we get $\Delta \simeq (p^2-\varPi_L)(p^2-\varPi_T-\varPi_\parallel)$.
Plugging this expression to Eq.~(\ref{eq11}), 
the gluon propagator near the mass-shell is obtained as 
\begin{align}
\begin{split}
D^{\mu \nu}_R
&\simeq -\frac{P^{\mu\nu}_T-P^{\mu\nu}_\perp}{p^2-\varPi_T-\varPi_\parallel}
-\frac{P^{\mu\nu}_\perp}{p^2-\varPi_T-\varPi_\perp}.
\end{split}
\end{align}
We note that the above two tensors are orthogonal, $(P^{\mu\nu}_T-P^{\mu\nu}_\perp)P^\perp_{\nu\alpha}=0$.

%%%%%%%%%%%%%%%%%%%%%%%%%%%%%%%%%%%%%%%%%%%%
\section{Properties of projection tensors}
\label{app:projection}

All the projection tensors are normalized as 
\begin{align}
P^{\mu\alpha}_i P^i_{\alpha\nu}
&= -P^{\mu}_{i \ \nu},
\end{align}
where $i=T, L, \parallel, \perp$.

The following pairs of the projection tensors are orthogonal: 
\begin{align}
P^{\mu\alpha}_T P^L_{\alpha\nu}
=P^{\mu\alpha}_\parallel P^\perp_{\alpha\nu}
=P^{\mu\alpha}_\perp P^L_{\alpha\nu}
=0.
\end{align}
On the other hand, multiplications for the other pairs result in 
nonvanishing structures as 
\begin{align} 
P^{\mu\alpha}_T P^\perp_{\alpha\nu}
&= P^{\mu\alpha}_\perp P^T_{\alpha\nu}
= -P^{\mu}_{\perp \nu} ,\\
\nonumber
P^{\mu\alpha}_T P^\parallel_{\alpha\nu}
&= -\frac{p^0}{p^2_\parallel \vp^2}
\left[-p^3p^\mu+p^0p^3n^\mu-\vp^2b^\mu \right] \\
&~~~\times \left[p^3n_\nu -p^0b_\nu\right].
\label{eq1}
\end{align}
We note that the right-hand side in Eq.~(\ref{eq1}) is not symmetric in $\mu$ and $\nu$.
However, when evaluating the inverse matrix to get the gluon propagator, 
one needs only a symmetrized combination 
\begin{eqnarray} 
&&  \hspace{-0.5cm}
P^{\mu\alpha}_T P^\parallel_{\alpha\nu}+P^{\mu\alpha}_\parallel P^T_{\alpha\nu}
\nonumber
\\
&&
= -\frac{1}{p^2_\parallel } \Bigl[ 2(p^0)^2  \left\{a n^\mu n_\nu+ b^\mu b_\nu \right\} 
\nonumber
\\
&& \hspace{1.5cm}
-p^0 p^3 (1+y^2)(b^\mu n_\nu+n^\mu b_\nu) 
\nonumber
\\
&& \hspace{1.5cm}
 -p^0 a (n^\mu p_\nu+p^\mu n_\nu)
+y^2 p^3 (b^\mu p_\nu+p^\mu b_\nu)
\Bigr] 
\nonumber
\\
&&
= \frac{1}{p^2_\parallel} \Bigl[ ap^2 P^{\mu}_{L \nu} 
+(p^0)^2(1-a) (P^{\mu}_{\perp \nu}-P^{\mu}_{T \nu})
\Bigr] 
 -P^{\mu}_{\parallel \nu}.
 \nonumber
 \\
\end{eqnarray}
By using a simple relation $P^{\mu\alpha}_L=P^{\mu\alpha}_0-P^{\mu\alpha}_T$, 
the product between $P^L$ and $P^\parallel$ can be obtained as 
$P^{\mu\alpha}_L P^\parallel_{\alpha\nu}+P^{\mu\alpha}_\parallel P^L_{\alpha\nu}
= -[P^{\mu\alpha}_T P^\parallel_{\alpha\nu}+P^{\mu\alpha}_\parallel P^T_{\alpha\nu}]-2P^{\mu}_{\parallel \nu}$, 
where the symmetrized tensor between the brackets are given just above.

In the Coulomb gauge, there appears another tensor $B^{\mu\nu}$ defined below Eq.~(\ref{eq2}).
Its product with the projection tensors are found to be 
\begin{align} 
B^{\mu\nu} P^\parallel_{\nu\alpha}
&= -\frac{(p^0)^2}{(p^3)^2}P^\mu_{\parallel \alpha}
-P^{\mu\nu}_T P^\parallel_{\nu\alpha}
+ \frac{(p^0)^2}{(p^3)^2} b^\mu P^\parallel_{3\alpha} 
,\\
B^{\mu\nu} P^T_{\nu\alpha}
&= \frac{p^0}{p^3} n^\mu P^T_{3\alpha} ,\\
\nonumber
B^{\mu\nu} P^L_{\nu\alpha}
&= 
\frac{1}{(p^3)^2}\Bigl[-p^2_L\left( P^\mu_{\parallel \alpha}+P^{\mu\nu}_\parallel P^T_{\nu\alpha}\right) 
+(p^0)^2b^\mu P^L_{3\alpha}\Bigr] \\
&~~~
-P^\mu_{L \alpha} ,\\
B^{\mu\nu} P^\perp_{\nu\alpha}
&= 0.
\end{align}

%%%%%%%%%%%%%%%%%%%%%%%%%%%%%%%%%%%%%%%%%%%%%%%%%%

\bibliography{reference}

\end{document}